\documentclass[11pt]{amsart}

\usepackage[USenglish]{babel}

\usepackage{amsmath,amsthm,amssymb,amscd}
\usepackage{booktabs}
\usepackage{url}

\usepackage{MnSymbol}

\usepackage{tikz}
\usetikzlibrary{arrows,shapes.geometric,positioning,decorations.markings}

\usepackage[mathscr]{eucal}
\usepackage[normalem]{ulem}
\usepackage{latexsym,youngtab}
\usepackage{multirow}
\usepackage{epsfig}
\usepackage{parskip}

\usepackage[margin=2cm]{geometry}
\usepackage{faktor}
\usepackage{xcolor}
\usepackage{geometry}
\newtheorem{innercustomthm}{Theorem}
\newenvironment{customthm}[1]
{\renewcommand\theinnercustomthm{#1}\innercustomthm}
{\endinnercustomthm}

\usepackage{color}

\newtheoremstyle{custom}
  {3pt}
  {3pt}
  {\slshape}
  {}
  {\bfseries}
  {.}
  { }
   {}
\theoremstyle{custom}
\newtheorem{theorem}{Theorem}[section]
\newtheorem{proposition}[theorem]{Proposition}
\newtheorem{observation}[theorem]{Observation}
\newtheorem{proposition/definition}[theorem]{Proposition/Definition}
\newtheorem{lemma}[theorem]{Lemma}

\theoremstyle{definition}
\newtheorem{definition}[theorem]{Definition}

\newtheorem{example}[theorem]{Example}

\newtheorem{question}[theorem]{Question}

\theoremstyle{remark}
\newtheorem{remark}[theorem]{Remark}

\newtheoremstyle{exercise}
  {3pt}
  {6pt}
  {}
  {}
  {\bfseries}
  {:}
  { }
   {}
\theoremstyle{exercise}
\newtheorem{exercise}[theorem]{Exercise}
\newtheoremstyle{exercises}
  {3pt}
  {6pt}
  {}
  {}
  {\bfseries}
  {:}
  {\newline}
   {}

\theoremstyle{exercise}
\newtheorem{exercises}[theorem]{Exercises}

\input epsf
\def\boxit#1{\vbox{\hrule height1pt\hbox{\vrule width1pt\kern3pt
  \vbox{\kern3pt#1\kern3pt}\kern3pt\vrule width1pt}\hrule height1pt}}

\def\trank{\text{rank}}

\def\BC{\mathbb C}

\def\BP{\mathbb P}
\def\pp#1{\mathbb P^{#1}}

\def\pp#1{{\mathbb P}^{#1}}
\def\tdim{{\rm dim}}

\def\hd{,...,}

\def\inv{{}^{-1}}

\def\cG{{\mathcal G}}
\def\cR{{\mathcal R}}

\def\11{\mathbf 1}

\def\fsl{{\mathfrak {sl}}}

\def\a{\alpha}

\def\b{\beta}
\def\g{\gamma}
\def\s{\sigma}

\def\ot{{\mathord{ \otimes } }}
\def\op{{\mathord{\,\oplus }\,}}

\def\ra{{\mathord{\;\rightarrow\;}}}

\def\La#1{\Lambda^{#1}}

\def\frak{\mathfrak}
\def\fsl{\frak s\frak l}

\def\op{\oplus}
\def\BZ{\Bbb Z}

\def\op{\oplus}

\def\ul{\underline}

\def\s{\sigma}

\def\a{\alpha}
\def\b{\beta}

\def\g{\gamma}

\def\FS{\mathfrak  S}

\def\ol{\overline}

\def\BP{\mathbb  P}
\def\BC{\mathbb  C}

\def\pp#1{\mathbb  P^{#1}}

\def\tcodim{\text{codim}}

\def\ci{\mathcal  I}

\def\hd{, \hdots ,}

\def\inv{{}^{-1}}

\def\La#1{\Lambda^{#1}}

\def\pp#1{\mathbb  P^{#1}}

\def\ur{\underline{\mathbf{R}}}

\def\ra{\rightarrow}

\def\tdeg{\operatorname{deg}}
\def\tdet{\operatorname{det}}

\def\tend{\operatorname{End}}

\def\tdim{\operatorname{dim}}

\def\tmin{\operatorname{min}}
\def\tmax{\operatorname{max}}

\def\trank{\operatorname{rank}}

\def\bbb{{\mathbf{b}}}

\def\be{\begin{equation}}
\def\ene{\end{equation}}
\def\aaa{{\mathbf{a}}}
\def\bbb{{\mathbf{b}}}
\def\ccc{{\mathbf{c}}}
\def\tsgn{{\rm{sgn}}}

\DeclareMathOperator\tspan{span}

\def\trank{\mathbf{R}}

\newcommand{\isom}{\cong}

\def\trker{{\rm Rker}}

\def\trker{{\rm Rker}}
\def\tspan{{\rm span}}

\newcommand{\Id}{\operatorname{Id}}

\def\Mn{M_{\langle \nnn \rangle}}\def\Mone{M_{\langle 1\rangle}}\def\Mtwo{M_{\langle 2 \rangle}}

\def\Mn{M_{\langle \nnn \rangle}}\def\Mone{M_{\langle 1\rangle}}

\def\Mtwo{M_{\langle 2\rangle}}

\def\trank{{\mathrm {rank}}}

\def\trker{{\rm Rker}}

\def\aaa{\mathbf{a}}
\def\bbb{\mathbf{b}}
\def\ccc{\mathbf{c}}

\def\trker{\operatorname{Rker}}

\def\Mn{M_{\langle n \rangle}}\def\Mone{M_{\langle
1\rangle}}

\def\Mtwo{M_{\langle 2\rangle}}

\def\trank{{\mathrm {rank}}}

\def\trker{{\rm Rker}}

\def\aaa{{\bold a}} \def\ccc{{\bold c}}

\def\uQ{\ul{\bold Q}}

\keywords{Matrix multiplication complexity, Tensor rank, Asymptotic rank, Laser method}


\renewcommand{\a}{\alpha}
\renewcommand{\b}{\beta}
\renewcommand{\g}{\gamma}

\renewcommand{\BC}{\mathbb{C}}

  \newtheorem{theorem/definition}{Theorem/Definition} 

\newcommand*\quot[2]{{^{\textstyle #1}\big/_{\textstyle #2}}}

\author[R. Geng, J.M. Landsberg]{Runshi Geng and J.M. Landsberg}

\address{Department of Mathematics, Texas 
A\&M University, College Station, TX 77843-3368, USA}
\email[R. Geng]{gengrunshi@math.tamu.edu}
\email[J. M. Landsberg]{jml@math.tamu.edu}
 
\title{On the geometry of geometric rank}

\thanks{Landsberg supported by NSF grant  AF-1814254}



\begin{document}
 \maketitle
 \begin{abstract}We make a geometric study of the {\it Geometric Rank} of tensors
 recently introduced by Kopparty et  al. Results include classification
 of tensors with degenerate geometric rank in $\BC^3\ot \BC^3\ot \BC^3$, classification
 of tensors with geometric rank two, and showing that upper bounds on geometric
 rank imply lower bounds on tensor rank.
 \end{abstract}
 
 \section{Introduction and statement of main results}
 Tensors appear in numerous mathematical and scientific contexts. The two contexts
 most relevant for this paper are quantum information theory and algebraic
 complexity theory, especially the study of the complexity of matrix
 multiplication.

 There are numerous notions of {\it rank} for tensors. One such, {\it analytic rank}, introduced in
 \cite{MR2773103} and developed further  in \cite{MR3964143}, is defined only over finite
 fields.
 In \cite{kopparty2020geometric} they define a new kind  of rank for tensors that
 is valid over arbitrary fields that is an asymptotic limit (as one enlarges the field)
 of analytic rank, that they call {\it geometric rank} (\lq\lq geometric\rq\rq\  in contrast to \lq\lq analytic\rq\rq ), and establish basic properties of geometric rank.
 In this paper we begin a systematic study of geometric rank and what it reveals about the
 geometry of tensors.
 
   Let  $T\in \BC^\aaa\ot \BC^\bbb\ot \BC^\ccc$ be a tensor and let
 $GR(T)$ denote the geometric rank of $T$ (see Proposition/Definition \ref{GRdef} below for the definition). For all tensors, one has  $GR(T)\leq \tmin\{\aaa,\bbb,\ccc\}$, and when $GR(T)<\tmin\{\aaa,\bbb,\ccc\}$, we say
 $T$ has {\it degenerate geometric rank}.  The case of geometric rank one was previously understood,
 see Remark \ref{GRone}.
 
 Informally, a tensor is
 {\it concise}  if it cannot be written as a tensor in a smaller ambient space  (see 
 Definition \ref{concisedef} below for the
 precise definition).

 Our main results are:
 \begin{itemize}
 \item Classification of    tensors  with   geometric rank two.
 In particular,  in $\BC^3\ot \BC^3\ot \BC^3$ there are exactly two concise tensors
 of geometric rank two, and in  $\BC^m\ot \BC^m\ot \BC^m$,      $m> 3$,  there is a unique 
 concise tensor with geometric rank two (Theorem
 \ref{GRtwo}).
 \item Concise $1_*$-generic tensors (see Definition \ref{1stardef})
 in $\BC^m\ot \BC^m\ot \BC^m$   with   geometric rank at most  three have tensor rank at least $2m-3$
and all other concise tensors of geometric rank at most three have
tensor rank at least $m+\lceil \frac {m-1}2 \rceil -2$  (Theorem \ref{GRthree}).  
 \end{itemize}
 
 We also  compute the geometric ranks of numerous tensors of interest
 in  \S\ref{Exsect},  and analyze the geometry associated to tensors with
 degenerate geometric rank in \S\ref{GRgeom}, where we also  point out especially intriguing
 properties of tensors in $\BC^m\ot \BC^m\ot \BC^m$  of minimal border rank.
 
 \subsection*{Acknowledgements} We thank Vladimir Lysikov for numerous
 comments and pointing out a gap in the proof of Theorem \ref{GRthree}
 in an earlier draft, Filip Rupniewski for pointing out
 an error in an earlier version of Proposition  \ref{comprex}, 
 Hang Huang for providing a more elegant proof of Lemma  \ref{linelem}
 than our original one,
 Harm Derksen for Remark \ref{gstablerem}, 
   and Fulvio Gesmundo,  Hang Huang, Christian Ikenemeyer and Vladimir Lysikov 
 for useful conversations. We also thank the anonymous referee for useful suggestions.
 
 \section{Definitions and Notation}

 Throughout this paper we give our vector spaces names: $A=\BC^\aaa,B=\BC^\bbb, C=\BC^\ccc$ 
 and we often will take $\aaa=\bbb=\ccc=m$. Write $\tend(A)$ for
 the space of linear maps $A\ra A$ and $GL(A)$ for the invertible linear maps.
 The dual space to $A$ is denoted $A^*$, its 
 associated projective space is $\BP A$, and for $a\in A\backslash 0$,
 we let $[a]\in \BP A$ be its projection to projective space.
 For a subspace $U\subset  A$, $U^\perp\subset A^*$ is its annihilator.
 For a subset $X\subset A$, $\langle X\rangle\subset A$ denotes its linear span. 
 We write $GL(A)\times GL(B)\times GL(C) \cdot T\subset A\ot B\ot C$ 
 for the orbit of $T$, and similarly for the images of $T$ under endomorphisms.
 For a set $X$, $\ol{X}$ denotes its closure in the Zariski topology (which, for
 all examples in this paper, will also be its closure in the Euclidean topology).
 
 Given $T\in A\ot B\ot C$, we let $T_A: A^*\ra B\ot C$ denote the corresponding
 linear map, and similarly for $T_B,T_C$.  We omit  the subscripts when there is no ambiguity. As examples, $T(A^*)$ means $T_A(A^*)$, and given $\beta\in B^*$, $T(\beta)$ means $T_B(\beta)$. 
 
   Fix bases $\{a_i\}$, $\{b_j\}$, $\{ c_k\}$ of $A,B,C$, let  $\{\a_i\}$, $\{\b_j\}$, $\{ \gamma_k\}$ be the corresponding dual bases of $A^*,B^*$ and $C^*$. The linear space $T(A^*)\subset B\otimes C$ is considered as a space of matrices, and is often presented as the image of a general point $\sum_{i =1}^\aaa x_i\a_i\in A^*$, i.e. a $\bbb\times \ccc$ matrix of linear forms in variables $\{x_i\}$. 
 
 Let $T\in A\ot B\ot C$. $T$ has {\it rank one} if
 there exists nonzero $a\in A$, $b\in B$, $c\in C$ such that $T=a\ot b\ot c$.
   
For $r\leq \tmin\{\aaa,\bbb,\ccc\}$,  write $\Mone^{\op r}=\sum_{\ell=1}^r a_\ell\ot b_\ell\ot c_\ell$.

We review   various notions of rank for tensors:
 \begin{definition} \
 \begin{enumerate} 
 \item The smallest $r$ such that $T$ is a sum of $r$ rank one
 tensors is called the {\it tensor rank} (or {\it rank}) of $T$ and is denoted $\bold R(T)$. This is  
 the smallest $r$ such that, allowing $T$ to be in a larger space,  $T\in \tend_r\times \tend_r\times\tend_r\cdot \Mone^{\op r}$.
 
 \item The smallest $r$ such that $T$ is a limit of rank $r$ tensors
 is called the {\it border rank} of $T$ and is denoted $\ur(T)$. This is  the smallest
 $r$ such that, allowing $T$ to be in a larger space,  $T\in \ol{GL_r\times GL_r\times GL_r\cdot \Mone^{\op r}}$.
 
 \item $(\bold{ml}_A,\bold{ml}_B,\bold{ml}_C):= (\trank\, T_A,\trank\, T_B,\trank\, T_C)$
 are the three {\it multi-linear ranks} of $T$.
 
 \item The largest  $r$ such that $\Mone^{\op r}\in \ol{GL(A)\times GL(B)\times GL(C)\cdot T}$
 is called the {\it border subrank}  $T$ and denoted $\ul{\bold Q}(T)$.

 \item The largest $r$ such that $\Mone^{\op r}\in \tend(A)\times \tend(B)\times \tend(C)\cdot T$ is called
 the {\it subrank} of $T$ and denoted $\bold Q(T)$.
 \end{enumerate}
 \end{definition}

 We have the inequalities
 $$\bold Q(T)\leq \uQ(T)\leq \tmin\{\bold{ml}_A,\bold{ml}_B,\bold{ml}_C\}
 \leq \tmax\{\bold{ml}_A,\bold{ml}_B,\bold{ml}_C\}\leq \ur(T)\leq \bold R(T),
 $$
 and all inequalities may be strict. 
 For example $\Mtwo$ of Example \ref{mmex} satisfies
 $\uQ(\Mtwo)=3$   \cite{kopparty2020geometric} and $\bold Q(\Mtwo)=2$ \cite[Prop. 15]{MR4210715}
 and all multilinear ranks are $4$.  Letting $\bbb\leq \ccc$, $T=a_1\ot (\sum_{j=1}^\bbb b_j\ot c_j)$
 has $\bold{ml}_A(T)=1$, $\bold{ml}_B(T)=\bbb$. 
 A generic tensor in $\BC^m\ot \BC^m\ot \BC^m$ satisfies $ \bold{ml}_A=\bold{ml}_B,=\bold{ml}_C=m$
 and $\ur(T)=O(m^2)$. The tensor 
 $T=a_1\ot b_1\ot c_2+a_1 \ot b_2\ot c_1+a_2\ot b_1\ot c_1$ satisfies
 $\ur(T)=2$ and $\bold R(T)=3$.
 We remark that  very recently Kopparty and Zuiddam (personal
 communication)  have shown that a generic
 tensor in $\BC^m\ot \BC^m\ot \BC^m$ has subrank at most $3m^{\frac 23}$. 
 
 In contrast,  the corresponding notions for matrices all coincide.
  
 \begin{definition} \label{concisedef}
 A tensor $T\in A\ot B\ot C$ is {\it concise} if
$\bold{ml}_A=\aaa$, $\bold{ml}_B=\bbb$, $\bold{ml}_C=\ccc$, 
 \end{definition}

The   rank and border rank of a tensor $T\in A\ot B\ot C$ measure the complexity of
evaluating the corresponding bilinear map $T: A^*\times B^* \ra C$ or trilinear form
$T: A^*\times B^*\times C^*\ra \BC$. A concise tensor in $\BC^m\ot \BC^m\ot \BC^m$ of rank $m$ (resp. border rank $m$),
is said to be of {\it minimal rank} (resp. {\it minimal border rank}). It is a longstanding
problem to characterize  tensors of minimal border rank, and how much larger
the rank can be than the border rank. The largest rank of any explicitly known 
sequence of tensors
is $3m-o(m)$ \cite{MR3025382}. While tests exist to bound the ranks of tensors,
previous to this paper there was no general geometric criteria that would 
lower bound tensor rank (now see Theorem \ref{GRthree} below). 
The border rank is measured by a classical geometric object: secant
varieties of Segre varieties. The border subrank, to our knowledge,
has no similar classical object. In this paper we discuss how
geometric rank is related to classically studied questions in algebraic
geometry:      linear spaces of matrices 
with large  intersections with   the variety of matrices
of rank at most $r$.  See  Equation \eqref{Xi} for a precise statement. 
 
 Another notion  of rank for tensors is the 
  {\it slice rank} \cite{Taoblog},  denoted by  $\mathrm{SR}(T)$: it  is the smallest $r$ such that there exist 
$r_1,r_2,r_3$ such that $r=r_1+r_2+r_3$, 
$A'\subset A$ of dimension $r_1$, $B'\subset B$ of dimension $r_2$, and
$C'\subset C$ of dimension $r_3$,  such that 
$T\in A'\ot B\ot C+ A\ot B'\ot C+ A\ot B\ot C'$.
It was originally introduced in the context of the cap set problem but has
turned out (in its asymptotic version) to be important for quantum information
theory and Strassen's laser method, more precisely, Strassen's theory
of asymptotic spectra, see \cite{MR3826254}.

\begin{remark}\label{gstablerem}
In \cite{derksen2020gstable} a notion of rank for tensors inspired by invariant theory,
called {\it $G$-stable rank} is introduced. Like geometric rank, it is bounded
above by the slice rank and below by the border subrank. Its relation to
geometric rank appears to be subtle:   the $G$-stable rank of
the matrix multiplication tensor
$\Mn$ equals   $n^2$, which is greater than the geometric rank (see Example \ref{mmex}), but
the $G$-stable rank of $W:=a_1\ot b_1\ot c_2+a_1\ot b_2\ot c_1+a_2\ot b_1\ot c_1$
is $1.5$ ($G$-stable rank need not be integer valued), while $GR(W)=2$.
\end{remark}

  Like multi-linear rank, 
geometric rank generalizes row rank and column
 rank of matrices, but unlike multi-linear rank, it salvages the fundamental theorem
 of linear algebra that row rank equals column rank.
 Let $Seg(\BP A^*\times \BP B^*)\subset \BP (A^*\ot B^*)$ denote the
 {\it Segre variety} of rank one elements.
 
  Let $\Sigma^{AB}_T = \{ ([\a],[\b])\in \BP A^*\times \BP B^* \mid T(\a,\b,\cdot)=0\}$, so
 \begin{align}
\label{sigmaab} Seg(\Sigma^{AB}_T)=  \BP (T(C^*)^{\perp})\cap Seg(\BP A^*\times \BP B^*)\end{align} 
and let  $\Sigma^A_j=\{[\a]\in \BP A^*\mid \trank(T(\a))\leq \tmin\{\bbb,\ccc\}-j\}$.
Let $\pi^{AB}_A: \BP A^*\times \BP B^* \ra   \BP A^*$ denote the projection.

\begin{proposition/definition}\cite{kopparty2020geometric}\label{GRdef}
The following quantites are all equal and called the {\it geometric rank} of $T$, denoted
$GR(T)$:
\begin{enumerate}
\item $\tcodim (\Sigma^{AB}_T, \BP A^*\times \BP B^*)$
\item $\tcodim (\Sigma^{AC}_T, \BP A^*\times \BP C^*)$
\item $\tcodim (\Sigma^{BC}_T, \BP B^*\times \BP C^*)$
\item $  \aaa+\tmin\{\bbb,\ccc\} -1 - \tmax_{j}(\tdim\Sigma^A_j +j) $
\item $  \bbb+\tmin\{\aaa,\ccc\}-1 - \tmax_{j}(\tdim\Sigma^B_j +j) $
\item $  \ccc+\tmin\{\aaa,\bbb\} -1 - \tmax_{j}(\tdim\Sigma^C_j +j) $.
\end{enumerate}
\end{proposition/definition}
\begin{proof}
The classical row rank equals column rank theorem implies that
when $\Sigma^A_j\neq \Sigma^A_{j+1}$,
 the   fibers of $\pi^{AB}_A$ are $\pp{j-1}$'s if $\bbb\geq \ccc$ and
 $\pp{j-1+\bbb-\ccc}$'s when $\bbb<\ccc$.
  The  variety  $\Sigma^{AB}_T$ 
is the union of the $(\pi^{AB}_A)\inv(\Sigma^A_j)$, which
have dimension $\tdim\Sigma^A_j+j-1$ when $\bbb\geq \ccc$ and $\tdim\Sigma^A_j+j-1+\bbb-\ccc$
when $\bbb<\ccc$.
The dimension of a variety is  the dimension  of a largest dimensional irreducible component.
\end{proof}
 
 \begin{remark} In \cite{kopparty2020geometric}
they  work with   $\hat\Sigma^{AB}_T:=\{ (\a,\b)\in A^*\times B^* \mid T(\a,\b,\cdot)=0\}$
and define 
 geometric rank to be  
 $ 
 GR(T):=\tcodim(\hat\Sigma^{AB}_T, A^*\times B^*)$. This is equivalent to our
 definition, which 
 is clear except  $0\times B^*$ and  $A^*\times 0$ are always contained in $\hat\Sigma^{AB}_T$ which implies $ {GR}(T)\leq \tmin\{\aaa,\bbb\}$ and by symmetry
 $GR(T)\leq \tmin\{\aaa,\bbb,\ccc\}$, but there is no corresponding set  in the projective variety $\Sigma^{AB}_T$. Since  \eqref{sigmaab} implies
 \begin{align*}
 \tdim \Sigma^{AB}_T&\geq 
   \tdim \BP (T(C^*)^\perp) + \tdim Seg(\BP A^*\times \BP B^*)- \tdim \BP (A^*\ot B^*)\\
 &= \aaa\bbb-\ccc-1+\aaa+\bbb-2-(\aaa\bbb-1)  \\
 &=\aaa+\bbb-\ccc-2
 \end{align*}
 we still have $ {GR}(T)\leq \aaa+\bbb-\ccc$ and by symmetry $GR(T)\leq \tmin\{\aaa,\bbb,\ccc\}$  using our definition.  We note that for tensors with more factors, one must be more careful when working
 projectively. 
 \end{remark}
 
 One has $\ul{\bold Q}(T)\leq GR(T)\leq SR(T)$ \cite{kopparty2020geometric}.
 In particular, one may use  geometric rank   to bound the border subrank. 
 An example of such a bound 
 was an important  application in \cite{kopparty2020geometric}.
 
 \begin{remark}\label{GRone}
 The set of tensors with slice rank one is the
 set of tensors living in some $\BC^1\ot B\ot C$
 (after possibly re-ordering and re-naming factors),  and the same is true for tensors with geometric rank one. Therefore for any tensor $T$, $GR(T)=1$ if and only if $\mathrm{SR}(T)=1$. 
 \end{remark}
 
 \begin{definition}\label{1stardef} Let $\aaa=\bbb=\ccc=m$.
  A tensor is {\it $1_A$-generic} if $T(A^*)\subset B\ot C$ contains
 an element of full rank $m$,  {\it binding} if it is at least two of  $1_A$, $1_B$, $1_C$ generic,
 {\it $1_*$-generic} if it is $1_A$, $1_B$ or $1_C$-generic, 
  and it is {\it $1$-generic} if it is
 $1_A,1_B$ and $1_C$-generic. A tensor is {\it $1_A$-degenerate} if it is not $1_A$-generic.
   Let $1_A-degen$ denote the variety 
 of tensors that are not $1_A$-generic,   
 and let  $1-degen$ the variety of tensors that are   $1_A,1_B$ and $1_C$ degenerate.
 \end{definition}
 
  $1_A$-genericity is important in the
 study of tensors as
 Strassen's equations \cite{Strassen505} and more generally
 Koszul flattenings \cite{MR3081636} fail
 to give good lower bounds for tensors that are $1$-degenerate.
 Binding tensors are those that arise as structure tensors of algebras, see \cite{MR3578455}.
 
 Defining equations for $1_A-degen$
 are given by the module $S^mA^*\ot \La m B^*\ot \La m C^*$, see \cite[Prop. 7.2.2.2]{MR2865915}.

\begin{definition} \label{bndrkdef} A subspace $E\subset B\ot C$ is of {\it bounded rank $r$}
 if for all $X\in E$, 
 $\trank(X)\leq r$. 
\end{definition}
 
\section{Statements of main results}
Let $\cG\cR_{s}(A\ot B\ot C)\subset \BP (A\ot B\ot C)$ denote
 the  set  of tensors of geometric rank at most $s$ which is Zariski closed \cite{kopparty2020geometric},
 and write $\cG\cR_{s,m}:=\cG\cR_{s}(\BC^m\ot \BC^m\ot \BC^m)$. 
 By Remark \ref{GRone}, $\cG\cR_{1}(A\ot B\ot C)$ is the variety  of tensors that
 live in some $\BC^1\ot B\ot C$, $A\ot \BC^1\ot C$,
 or $A\ot B\ot \BC^1$.
 
 In what follows, a statement of the form \lq\lq there exists a unique tensor...\rq\rq,
 or \lq\lq there are exactly two tensors...\rq\rq,
 means    up to the action of $GL(A)\times GL(B)\times GL(C)\rtimes \FS_3$.

\begin{theorem} \label{GRtwo} For $\aaa,\bbb,\ccc\geq3$, 
the variety $\mathcal{GR}_{2}(A\ot B\ot C)$ is the variety of tensors $T$ such that $T(A^*)$, $T(B^*)$,
 or $T(C^*)$ has bounded rank 2. 
 
There are exactly two concise tensors in $\mathbb{C}^3\otimes\mathbb{C}^3\otimes\mathbb{C}^3$ with $GR(T)=2$:
\begin{enumerate}
\item The unique up to scale skew-symmetric tensor $T=\sum_{\s\in \FS_3}\tsgn(\s) a_{\s(1)}\ot b_{\s(2)}\ot c_{\s(3)}\in \La 3\BC^3\subset  \BC^3\ot \BC^3\ot \BC^3$

and

\item $T_{utriv,3}:=a_1\ot b_1\ot c_1 + a_1\ot b_2\ot c_2+ a_1\ot b_3\ot c_3
+a_2\ot b_1\ot c_2+a_3\ot b_1\ot c_3\in S^2\BC^3\ot \BC^3\subset  \BC^3\ot \BC^3\ot \BC^3$. 
\end{enumerate}
There is a unique  concise  tensor $T\in  \BC^m\ot \BC^m\ot \BC^m$
satsifying $GR(T)=2$ when $m>3$,    
  namely
 $$
 T_{utriv,m}:=
 a_1\ot b_1\ot c_1+   \sum_{\rho=2}^m [a_1\ot  b_\rho\ot c_\rho +   a_\rho\ot b_1\ot c_\rho].
 $$
 This tensor satisfies $\ur(T_{utriv,m})=m$ and $\bold R(T_{utriv,m})=2m-1$. 
\end{theorem}

In the $m=3$ case (1) of Theorem \ref{GRtwo} we have $\Sigma^{AB}_T\isom \Sigma^{AC}_T\isom \Sigma^{BC}_T\isom \BP A^*\subset \BP A^*\times \BP A^*$
embedded diagonally and $\Sigma^A_1=\Sigma^B_1=\Sigma^C_1=\BP A^*$.

In the $m=3$ case (2) of Theorem \ref{GRtwo}  we have 
\begin{align*}
\Sigma^{AB}_T&=\BP\langle \a_2,\a_3\rangle \times \BP \langle \b_2,\b_3\rangle=
\pp 1\times \pp 1\\
\Sigma^{AC}_T&=\{([ s\a_2+t\a_3 ] , [
u\g_1 + v( -t\g_2+s\g_3)])\in \mathbb{P}A\times\mathbb{P}C  \mid [s,t]\in \pp 1, [u,v]\in \pp 1\}  \\
\Sigma^{BC}_T&=\{([ s\b_2+t\b_3 ], [
u\g_1 + v( -t\g_2+s\g_3)])\in\mathbb{P}B\times\mathbb{P}C  \mid [s,t]\in \pp 1, [u,v]\in \pp 1  \}.
\end{align*}
 If one looks at the scheme structure,
  $\Sigma^A_2$, $\Sigma^B_2$ are lines with multiplicity three and $\Sigma^C_1=\BP C^*$.

 \begin{remark}
 The tensor $T_{utriv,m}$ has appeared several times in the literature:
 it is the structure tensor of the trivial algebra with unit (hence the name), and
 it has the largest symmetry group of any binding tensor \cite[Prop. 3.2]{2019arXiv190909518C}.
   It is also closely related to Strassen's tensor of \cite{MR882307}: it
 is the sum of Strassen's tensor with a rank one tensor.
 \end{remark}

Theorem \ref{GRtwo} is proved in \S\ref{GRtwopf}.

\begin{theorem} \label{GRthree}
Let $T\in A\ot B\ot C$ be concise and assume $\ccc\geq\bbb\geq \aaa>4$.
If $GR(T)\leq 3$, then  
$\bold R(T)\geq  \bbb+ \lceil {\frac {\aaa-1}2}\rceil -2$.

If moreover $\aaa=\bbb=\ccc=m$ and $T$ is 
  $1_*$-generic, then $\bold R(T)\geq 2m-3$.
\end{theorem}

In contrast to $\cG\cR_{1,m}$ and $\cG\cR_{2,m}$, the variety $\cG\cR_{3,m}$
is not just the  the set of tensors $T$ such that $T(A^*),T(B^*)$ or $T(C^*)$ has bounded rank 3. Other examples include
the structure tensor for $2\times 2$ matrix multiplication $\Mtwo$ (see Example \ref{mmex}),
the large and small Coppersmith-Winograd tensors (see Examples
\ref{CWqex} and \ref{cwqex}) and others (see \S\ref{moregr3}).

 Theorem \ref{GRthree}  gives the first algebraic way to lower bound tensor rank. Previously,
 the only technique to bound tensor rank beyond border rank was the {\it substitution
 method} (see \S\ref{submethrev}), which is not algebraic or systematically implementable.
 
Theorem \ref{GRthree} is proved in \S\ref{GRthreepf}.

 \section{Remarks on the geometry of geometric rank}\label{GRgeom}

 \subsection{Varieties arising in the study of geometric rank}
 Let $G(m,V)$ denote the Grassmannian of $m$-planes through the origin in the
 vector space $V$. 
 Recall the correspondence (see, e.g., \cite{MR3682743}):
 \begin{align}\nonumber 
 &\{ A\text{-concise tensors}\ T\in A\ot B\ot 
 C\}/\{ GL(A)\times GL(B)\times GL(C)-{\rm equivalence}\}\\
&\label{tscorresp} \leftrightarrow\\
&\nonumber  \{ \aaa-{\rm planes} \ E\in G(\aaa , B\ot C)\}/\{  GL(B)\times GL(C)-{\rm equivalence}.\}
\end{align}
 
It makes sense to study the $\Sigma^A_j$
 separately, as they have different geometry.
 To this end define  $GR_{A,j}(T)= \aaa+\tmin\{\bbb,\ccc\}-1 -  \tdim\Sigma^A_j -j$.
 Let  $\cG\cR_{r,  A,j}(A\ot B\ot C)=\{[T]\in \BP (A\ot B\ot C ) \mid   GR_{A,j}(T)\leq r\}$.
 
 Let $\s_r(Seg(\BP B\times \BP C))\subset \BP (B\ot C)$ denote
 the variety of $\bbb\times \ccc$ matrices of rank at most $r$. 
   
 By the correspondence \eqref{tscorresp}, the study of $\cG\cR_{r,A,j}(A\ot B\ot C)$ is the
 study of the variety    
\be\label{Xi}
  \{ E\in G(\aaa ,B^*\ot C^*)
 \mid \tdim(\BP E\cap \s_{\tmin\{\bbb,\ccc\}-j}(\BP B^*\times \BP C^*))\geq \aaa+\tmin\{\bbb,\ccc\} -j-1-r\}.
\ene

The following is immediate from the definitions, but since it is significant
we record it: 

 \begin{observation} $\cG\cR_{\aaa-1,  A,1}(A\ot B\ot C) = 1_A-degen$. In particular, tensors that are
 $1_A$, $1_B$, or $1_C$ degenerate  have degenerate geometric rank.
 
 $\cG\cR_{\aaa-1 , A,\aaa}(A\ot B\ot C)$ is the set of tensors that fail to be $A$ concise. In particular,
 non-concise tensors do not have maximal geometric rank.
 \end{observation}

It is classical  that  $\tdim \s_{m-j}(Seg(\pp{m-1}\times \pp{m-1}))=m^2-j^2-1$.
Thus for a general tensor in $\BC^m\ot \BC^m\ot \BC^m$, $\tdim(\Sigma^A_j)=m-j^2$. In particular, it 
is   empty when $j>\sqrt{m}$.

 \begin{observation}\label{Rmin} If $T\in\BC^m\ot \BC^m\ot \BC^m$ is concise and $GR(T)<m$, then
 $\bold R(T)>m$.
 \end{observation}
 \begin{proof} If $T$ is concise $\bold R(T)\geq m$, and for
 equality to hold  it can be written as $\sum_{j=1}^m a_j\ot b_j\ot c_j$ for some bases $\{a_j\},\{b_j\}$ and $\{c_j\}$ of $A,B$ and $C$ respectively.  But $GR(\sum_{j=1}^m a_j\ot b_j\ot c_j)=m$.
 \end{proof}

 \begin{question} For concise  $1_*$-generic tensors $T\in \BC^m\ot \BC^m\ot\BC^m$,
 is $\bold R(T)\geq 2m-GR(T)$?
 \end{question}

 \subsection{Tensors of minimal border rank}
 
If $T\in A\ot B\ot C=\BC^m\ot \BC^m\ot \BC^m$ is concise of minimal border rank $m$,
then there exist complete flags  in $A^*,B^*,C^*$, $0\subset  A_1^*\subset  A_2^*\subset 
\cdots \subset  A_{m-1}^*\subset  A^*$ etc..
such that $T|_{A_j^*\ot B_j^*\ot C_j^*}$ has border rank at most $j$, see \cite[Prop. 2.4]{CHLlaser}.
In particular, $\tdim( \BP T(A^*)\cap \s_j(\BP B\times \BP C))\geq j-1$.
If the inequality is strict for some $j$, say equal to $j-1+q$, we say   {\it the $(A,j)$-th flag condition
for minimal border rank is passed with   excess $q$}.

\begin{observation} The geometric rank of
a concise tensor in $\BC^m\ot \BC^m\ot \BC^m$
 is $m$ minus the largest excess of the $(A,j)$ flag conditions for minimal
border rank.
\end{observation}

We emphasize that a tensor with degenerate geometric rank need not have minimal
border rank, and need not   pass all the $A$-flag conditions for minimal border rank,
just that one of the conditions is passed with excess.

 \section{Examples of tensors with degenerate geometric ranks}\label{Exsect}

 \subsection{Matrix multiplication and related tensors}
 
\begin{example}[Matrix multiplication]\label{mmex} Set $m=n^2$. Let $U,V,W=\BC^n$.
Write $A=U^*\ot V$, $B=V^*\ot W$, $C=W^*\ot U$.
The structure tensor of matrix multiplication is 
$T=\Mn=\Id_U\ot\Id_V\ot \Id_W$ (re-ordered), where $\Id_U\in U^*\ot U$ is
the identity.

When $n=2$,  $\Sigma^{AB}=Seg(\BP U^*\times \ci_V\times \BP W)$,
where $\ci_V=\{ [v]\times [\nu]\in \BP V\times \BP V^* \mid \nu(v)=0\}$ has dimension $3$, so $GR(\Mtwo)=6-3=3$.
Note that  
$\Sigma^A_2=
\Sigma^A_1=Seg(\BP U^*\times \BP V)=Seg(\pp 1\times\pp 1)$ (with multplicity two).
For $[\mu\ot v]\in \Sigma^A_2$, $\pi_A^{-1}[\mu\ot v]=
\BP (\mu\ot v\ot v^\perp\ot W)\isom \pp 1$. Since
the tensor is $\BZ_3$-invariant the same holds for $\Sigma^B,\Sigma^C$.

For larger $n$,  the dimension of the fibers of $\pi^{AB}_A$ varies with the
rank of $X\in \{\tdet_{n}=0\}$. The fiber is
$[X]\times \BP (\trker(X)\ot W)$, which has dimension  $(n-\trank(X))n-1$.
Write $r=\trank(X)$.
Each $r$  gives rise to a $(n-r)n-1+(2nr-r^2-1)=n^2-r^2+nr-2$ dimensional component
of $\Sigma^{AB}$. There are  $n-1$ components, the largest dimension 
  is attained when $r=\lceil \frac n2\rceil$, where the
  dimension is  $n^2-\lceil \frac n2\rceil\lfloor \frac n2\rfloor -2$
and we recover the result of  \cite{kopparty2020geometric} that $GR(\Mn)=\lceil \frac 34 n^2\rceil=\lceil \frac 34 m\rceil$, caused by $\Sigma^A_{\lceil \frac n 2\rceil}$.
\end{example}

\begin{example}[Structure Tensor of $\mathfrak{sl}_n$] Set $m=n^2-1$.
Let $U=\BC^n$ , let $A=B=C=\fsl_n=\fsl(U)$. For $a,b\in \fsl_n$, $[a,b]$ denotes their commutator. 
Let $T_{\fsl_n}\in\mathfrak{sl}_n(\mathbb{C})^{\otimes 3}$ be 
the structure tensor of $\fsl_n$:  $T_{\fsl_n}=\sum_{ij=1}^{n^2-1} a_i\otimes b_j\otimes [a_i,b_j]$. Then $\hat{\Sigma}^{AB} =\{(x,y)\in A^*\times B^*|[x,y]=0\}$.

Let   $C(2,n):=\{(x,y)\in U^*\ot U\times U^*\ot U\,|\,xy=yx\}$. In \cite{MR86781} it was shown that
$C(2,n)$ is irreducible. Its dimension is
 $n^2+n$, which 
 was computed in  \cite[Prop. 6]{MR1753173}.
Therefore
  $\hat{\Sigma}^{AB}=(\mathfrak{sl}_n(\mathbb{C})\times\mathfrak{sl}_n(\mathbb{C}))\cap C(2,n)$
  has dimension $n^2+n-2$, and $GR(T_{\fsl_n})=\mathrm{dim}(\mathfrak{sl}_n(\mathbb{C})\times\mathfrak{sl}_n(\mathbb{C}))-\mathrm{dim}\hat{\Sigma}^{AB}=n^2-n
=m-\sqrt{m+1}$.
\end{example}



\begin{example}[Symmetrized Matrix Multiplication] Set $m=n^2$.
Let $A=B=C=U^*\ot U$, with $\tdim U=n$. 
Let $T=SM_{<n>}\in (U^*\ot U)^{\otimes 3}$ be 
the symmetrized matrix multiplication tensor:  $SM_{<n>}(X,Y,Z):=\mathrm{tr}(XYZ)+\mathrm{tr}(YXZ)$.
In \cite{MR3829726} it was shown that the exponent of $SM_{\langle n\rangle}$ equals the exponent
of matrix multiplication. On the other hand, $SM_{\langle n\rangle}$ is a cubic polynomial and thus
may be studied with more tools from classical algebraic geometry, which raises the hope of new 
paths towards determining the exponent. 
 Note that $SM_{<n>}(X,Y,\cdot)=0$ if and only if $XY+YX=0$. So $\hat{\Sigma}^{AB}=\{(X,Y)\in U^*\ot U\times U^*\ot U\,|\,XY+YX=0\}$.

Fix any matrix $X$, let $M_X$ and $M_{-X}$ be two copies of $\mathbb{C}^n$ with $\mathbb{C}[t]$-module structures: $t\cdot v:=Xv,\forall v\in M_X$ and $t\cdot w:=-Xw,\forall w\in M_{-X}$, where $\mathbb{C}[t]$ is the polynomial ring.

For any linear map $\varphi:M_X\rightarrow M_{-X}$,
\begin{align*}
\varphi\in\mathrm{Hom}_{\mathbb{C}[t]}(M_X,M_{-X}) & \iff\varphi(tv)=t\varphi(v),\forall v\in M_X\\
&\iff\varphi(Xv)=-X\varphi(v), \forall v\in M_X\\
&\iff\varphi X=-X \varphi.
\end{align*}
This gives a vector space isomorphism $(\pi^{AB}_A)^{-1}(X):=\{Y|XY+YX=0\}\cong \mathrm{Hom}_{\mathbb{C}[t]}(M_X,M_{-X})$.

By the structure theorem of finitely generated modules over
principal ideal domains, $M_X$ has a primary decomposition:
$$M_X\cong \quot{\mathbb{C}[t]}{(t-\lambda_1)^{r_1}}\oplus\cdots\oplus\quot{\mathbb{C}[t]}{(t-\lambda_k)^{r_k}}$$
for some $\lambda_i\in\mathbb{C}$ and $\sum r_i=n$. Replacing $t$ with $-t$ we get a decomposition of $M_{-X}$:
$$M_{-X}\cong \quot{\mathbb{C}[t]}{(t+\lambda_1)^{r_1}}\oplus\cdots\oplus\quot{\mathbb{C}[t]}{(t+\lambda_k)^{r_k}}.
$$

We have the decomposition  $\mathrm{Hom}_{\mathbb{C}[t]}(M_X,M_{-X})\cong \bigoplus\limits_{i,j}\mathrm{Hom}_{\mathbb{C}[t]}(\quot{\mathbb{C}[t]}{(t-\lambda_i)^{r_i}},\quot{\mathbb{C}[t]}{(t+\lambda_j)^{r_j}})$. For each $i,j$:
\begin{align*}&\mathrm{Hom}_{\mathbb{C}[t]}\left(\quot{\mathbb{C}[t]}{(t-\lambda_i)^{r_i}},\quot{\mathbb{C}[t]}{(t+\lambda_j)^{r_j}}\right)\\
&=\left\{
                \begin{array}{ll}
               \langle 1\mapsto(t-\lambda_i)^l\,|\,0\leq l\leq r_j-1\rangle &\mathrm{if}\; \lambda_i+\lambda_j=0\;\mathrm{and}\;r_i\geq r_j;\\
                \langle 1\mapsto(t-\lambda_i)^l\,|\,r_j-r_i\leq l\leq r_j-1\rangle &\mathrm{if}\; \lambda_i+\lambda_j=0\;\mathrm{and}\;r_i< r_j;\\
                0&\mathrm{otherwise.}
                \end{array}
                \right.
\end{align*}
Let $d_{ij}(X)$
denote its dimension,  then $d_{ij}(X)=\left\{
                \begin{array}{ll}
                \mathrm{min}\{r_i,r_j\}&\mathrm{if}\; \lambda_i+\lambda_j=0;\\
                0&\mathrm{otherwise.}
                \end{array}
                \right.$
Thus  $\mathrm{dim}((\pi^{AB}_A)^{-1} (X))=\sum\limits_{i,j} d_{ij}(X)$.

Each direct summand $\quot{\mathbb{C}[t]}{(t-\lambda_i)^{r_i}}$ of $M_X$ corresponds to a Jordan block of the Jordan canonical form of $X$ with size $r_i$ and eigenvalue $\lambda_i$, denoted as $J_{\lambda_i}(r_i)$.

Assume $X$ has eigenvalues $\pm\lambda_1, \cdots,\pm\lambda_k,\lambda_{k+1},\cdots,\lambda_{l}$ such that $\lambda_i\neq\pm\lambda_j$ whenever $i\neq j$. Let $q_{X,1}(\lambda)\geq q_{X,2}(\lambda) \geq \cdots$ be the decreasing sequence of sizes of the Jordan blocks of $X$ corresponding to the eigenvalue $\lambda$. Let $W(X)$ be the set of matrices $X'$ with eigenvalues $\pm\lambda'_1, \cdots,\pm\lambda'_k,\lambda'_{k+1},\cdots,\lambda'_{l}$ such that $\lambda'_i\neq\pm\lambda'_j$ whenever $i\neq j$, and $q_{X,j}(\pm \lambda_i)=q_{X',j}(\pm \lambda'_i)\,\forall i,j$. Then $W(X)$ is quasi-projective and irreducible  of dimension $\mathrm{dim}W(X)=\mathrm{dim}\{P^{-1}XP\,|\,\mathrm{det}P\neq 0\}+l$, and $(\pi^{AB}_A)^{-1}(X')$ is
of the same dimension as $(\pi^{AB}_A)^{-1}(X)$ for all $X'\in W(X)$.

By results in \cite{MR1355688},   the codimension of the   orbit of $X$ 
under the adjoint action of $GL(U)$ is $c_{Jor}(X):=\sum_{\lambda}[q_{X,1}(\lambda)+3q_{X,2}(\lambda)+5q_{X,3}(\lambda)+\cdots]$. Then 
 $$\mathrm{dim}\hat{\Sigma}^{AB}=\max_X(\mathrm{dim}W(X)+\mathrm{dim}\pi^{-1}_1(X))=\max_X(n^2-c_{Jor}(X)+ \tdim(\pi_{A}^{AB})\inv(X)+l)$$
because  
  $\hat \Sigma^{AB}=\cup_X(\pi^{AB}_A)\inv (W(X))$ is a finite union.

It is easy to show that $\tdim(\pi_{A}^{AB})\inv(X)-c_{Jor}(X)$ takes maximum $0$ when for every eigenvalue $\lambda_i$ of $X$, $-\lambda_i$ is an eigenvalue of $X$ and $q_{X,j}(\lambda_i)=q_{X,j}(-\lambda_i),\forall i,j$. So the total maximum is achieved when $X$ has the maximum possible number of  distinct pairs $\pm\lambda_i$, i.e.,  
$$X\cong\left\{ 
                \begin{array}{ll}
                \mathrm{diag}(\lambda_1,-\lambda_1,\lambda_2,-\lambda_2,\cdots,\lambda_{\frac{n}{2}},-\lambda_{\frac{n}{2}}) &\mathrm{if}\; n\; \mathrm{is\;even};\\
                \mathrm{diag}(\lambda_1,-\lambda_1,\lambda_2,-\lambda_2,\cdots,\lambda_{\frac{n-1}{2}},-\lambda_{\frac{n-1}{2}},0) &\mathrm{if}\; n\; \mathrm{is\;odd}.
                \end{array}
                \right.
$$

In both cases  $\mathrm{dim}\hat{\Sigma}^{AB}=n^2+\lfloor \frac{n}{2}\rfloor$. We conclude that $GR(s\Mn)=n^2-\lfloor \frac{n}{2}\rfloor=m-\lfloor \frac {\sqrt{m}}2\rfloor$. 
\end{example}

\subsection{Large border rank and small geometric rank}
The following example  shows  that border rank can be quite large while
 geometric rank is small:
 
 \begin{example}
Let $T\in A\ot B\ot C=\BC^m\ot \BC^m\ot \BC^m$
have the form
$T=a_1\ot (b_1\ot c_1+\cdots + b_m\ot c_m)+ T'$
where $T'\in A'\ot B'\ot C':=\tspan\{ a_2\hd a_m\}
\ot \tspan\{ b_1\hd b_{\lfloor \frac m2\rfloor}\}
\ot \tspan\{ c_{\lceil \frac m2\rceil}\hd c_m\}$, where $T'$ is generic. 
It was shown in \cite{MR3682743} that  $\bold R(T)= \bold R(T')+m$ and
$\ur(T)\geq \frac{m^2}8$.
 
We have
$$T(A^*)\subset \begin{pmatrix}
x_1 & & & & & \\
& \ddots & & & & \\
& & x_1 & & & &\\
* & \cdots & * & x_1 & &\\
\vdots & \vdots & \vdots & & \ddots & \\
* &\cdots & * & & & x_1\end{pmatrix}.
$$

Setting $x_1=0$, we see
a component of  $\Sigma^A_{\lfloor \frac m2\rfloor}\subset \BP A^*$ is a hyperplane
so $GR(T)\leq \lceil \frac m2\rceil+1$.
\end{example}

\subsection{Tensors arising in Strassen's laser method}

\begin{example}[Big Coppersmith-Winograd tensor] \label{CWqex}
The following tensor has been used to obtain every new upper bound on the 
exponent of matrix multiplication since 1988:
$$T_{CW,q}=\sum_{j=1}^q a_0\ot b_j\ot c_j
+a_j\ot b_0\ot c_j+a_j\ot b_j\ot c_0+a_0\ot b_0\ot c_{q+1}
+a_0\ot b_{q+1}\ot c_0+a_{q+1}\ot b_0\ot c_0.
$$
One has  
  $\bold R(T_{CW,q})=2q+3=2m-1$ \cite[Prop. 7.1]{MR3682743}
  and  $\ur(T_{CW,q})=q+2=m$  \cite{MR91i:68058}.
Note
$$
T_{CW,q}(A^*)=
\begin{pmatrix}
x_{q+1}& x_1 &\cdots & x_q &x_0\\
x_1 & x_0 & &  & \\
x_2 & & \ddots & & \\
\vdots & &  &   & \\
x_q & & & x_0 & \\
x_0 & & & & 0 
\end{pmatrix}
\ \ \simeq \ \ 
\begin{pmatrix}
x_0& x_1 &\cdots & x_q &x_{q+1}\\
 & x_0 & &  & x_1\\
 & & \ddots & & x_2\\
 & &  &   & \vdots\\
 & & & x_0 & x_q\\
 & & & & x_0
\end{pmatrix}
$$
where $\simeq$ means equal up to changes of bases. So we have 
   $\Sigma^A_{1}=\Sigma^A_2=\cdots =\Sigma^A_{q}=\{x_0=0\}$ and $\Sigma^A_{q+1}=\{x_0=\cdots=x_q=0\}$.
Therefore $GR(T_{CW,q})=2(q+2)-1-(\mathrm{dim}\Sigma^A_{q}+q)=3$.
\end{example}

\begin{example}[Small Coppersmith-Winograd tensor]\label{cwqex} The following tensor was
the second tensor used in the laser method and for $2\leq q\leq 10$, it
could potentially prove the exponent is less than $2.3$:
$T_{cw,q}=\sum_{j=1}^q a_0\ot b_j\ot c_j
+a_j\ot b_0\ot c_j+a_j\ot b_j\ot c_0$. 
It satisfies 
  $\bold R(T_{cw,q})=2q+1=2m-1$ \cite[Prop. 7.1]{MR3682743}
  and   $\ur(T_{cw,q})=q+2=m+1$  \cite{MR91i:68058}. We again have 
$GR(T_{cw,q})=3$ as e.g., $\Sigma^{AB}=\{ x_0=y_0=\sum_{j\geq 1} x_j y_j=0\}\cup\{\forall j\geq 1, x_j=y_j=0\}$.
\end{example}
  
  \begin{example} [Strassen's tensor] \label{strassenten} The following is the 
first tensor that was used in the laser method:
  $T_{str,q}=\sum_{j=1}^q a_0\ot b_j\ot
  c_j + a_j\ot b_0\ot c_j\in \BC^{q+1}\ot \BC^{q+1}\ot \BC^q$.
It satisfies $\ur(T_{str,q})=q+1$ and   $\bold R(T_{str,q})=2q$ \cite{MR3682743}.
Since
$$
T_{str,q}(A^*)=
\begin{pmatrix}  x_1& \cdots & x_q\\
  x_0 &  &  \\
&   \ddots  & \\
& &   x_0\end{pmatrix}
$$
We see $GR(T_{str,q})=2$ caused by $\Sigma^{A}_q=\BP \langle\a_1\hd \a_q\rangle$.
       \end{example}
       
 \subsection{Additional  examples of tensors  with geometric
 rank $3$}\label{moregr3}      
       
 \begin{example} \label{1ggr3} The following tensor
 was shown in \cite{MR3682743}  to take minimal values for Strassen's functional (called maximal
 compressibility in \cite{MR3682743}):
 $$
 T_{maxsymcompr,m}
 =a_1\ot b_1\ot c_1
 +\sum_{\rho=2}^m a_1\ot b_\rho\ot c_\rho
 + a_\rho\ot b_1\ot c_\rho + a_\rho\ot b_\rho\ot c_1.
 $$
 Note 
 $$T_{maxsymcompr,m}(A^*)=\begin{pmatrix}
 x_1 & x_2& x_3& \cdots & x_m\\
 x_2 & x_1&0&\cdots&0 \\
 x_3 & 0 & x_1 &&\vdots \\
 \vdots & \vdots & & \ddots &0\\
 x_m& 0&\cdots &0& x_1
 \end{pmatrix}.
 $$
Restrict to the hyperplane $\a_1=0$, we obtain 
 a space of bounded rank two, i.e.,  $\Sigma^A_{m-2}\subset \BP A^*$
 is a hyperplane. We conclude, assuming $m\geq 3$, that
  $GR(T)= 3$.
  \end{example}
  
  \begin{example}\label{badex}
  Let $m=2q$ and let
  $$
T_{gr3,1deg,2q}:=\sum_{s=1}^{q} a_s\ot b_1\ot c_s+ \sum_{t=2}^{q}a_{t+q-1}\ot b_t\ot c_1
+a_m\ot (\sum_{u=q+1}^m b_u\ot c_u),
$$
so 
\be\label{badA}
T_{gr3,1deg,m}(A^*)=
\begin{pmatrix} x_1 & x_2 & \cdots & x_q & 0 & \cdots & 0\\
x_{q+1}& 0 & \cdots & 0 &0 &\cdots & 0 \\
\vdots & \vdots & &\vdots &\vdots & &\vdots \\
x_{m-1}& 0 & \cdots &0 &0 &\cdots&0 \\
0 &0 &\cdots  & 0 &x_m &  & \\
\vdots &\vdots & & \vdots & &\ddots & \\
0 &0 &\cdots  & 0 &  &  &x_m \end{pmatrix}.
\ene
Then $GR(T_{gr3,1deg,m})=3$ (set $x_m=0$)  and $\bold R(T)=\frac 32m-1$,
the upper bound is clear from the expression the lower bound is given in
Example \ref{badgrex}.
\end{example}

  \begin{example}\label{badex2}
  Let $m=2q-1$ and let
  $$
T_{gr3,1deg,2q-1}:=\sum_{s=2}^{q} a_s\ot b_1\ot c_s+ a_{s+q-1}\ot b_s\ot c_1
+a_1\ot (\sum_{u=q+1}^m b_u\ot c_u),
$$
so 
\be\label{badA2}
T_{gr3,1deg,2q-1}(A^*)=
\begin{pmatrix} 0 & x_2 & \cdots & x_q & 0 & \cdots & 0\\
x_{q+1}& 0 & \cdots & 0 &0 &\cdots & 0 \\
\vdots & \vdots & &\vdots &\vdots & &\vdots \\
x_{m}& 0 & \cdots &0 &0 &\cdots&0 \\
0 &0 &\cdots  & 0 &x_1 &  & \\
\vdots &\vdots & & \vdots & &\ddots & \\
0 &0 &\cdots  & 0 &  &  &x_1\end{pmatrix}.
\ene
Then $GR(T_{gr3,1deg,2q-1})=3$ (set $x_1=0$)  and $\bold R(T_{gr3,1deg,2q-1})=  m+\frac{m-1}2-2$,
the upper bound is clear from the expression and  the lower bound is given in
Example \ref{badgrex}.
\end{example}

\subsection{Kronecker powers of tensors with degenerate geometric rank}
For tensors $T\in A\ot B\ot C$ and $T'\in A'\ot B'\ot C'$, the {\it Kronecker product} of $T$ and $T'$ is the tensor $T\boxtimes T' := T \ot T' \in (A\ot A')\ot (B\ot B')\ot (C\ot C')$, regarded as a $3$-way tensor. Given $T \in A \otimes B \otimes C$, the {\it Kronecker powers}  of $T$ are $T^{\boxtimes N} \in A^{\otimes N} \otimes B^{\otimes N} \otimes C^{\otimes N}$, defined iteratively. Rank and border rank are submultiplicative under the Kronecker product, while
subrank and border subrank are super-multiplicative under the Kronecker product.
 
Geometric rank is neither sub- nor  super-multiplicative under the Kronecker product.
We already saw the lack of sub-multiplicativity with $\Mn$ (recall $\Mn^{\boxtimes 2}=
M_{\langle n^2\rangle}$):
$\frac 34n^2=GR(\Mn^{\boxtimes 2})> \frac{9}{16}n^2=GR(\Mn)^2$.
An indirect example of the failure
of super-multiplicativity  is given in \cite{kopparty2020geometric} where they point out that some power of 
$W:=a_1\ot b_1\ot c_2+a_1\ot b_2\ot c_1+a_2\ot b_1\ot c_1$ is 
strictly sub-multiplicative. We make this explicit:

\begin{example}
 With basis indices ordered $22,21,12,11$ for $B^{\ot 2},C^{\ot 2}$, we have
$$
 W^{\boxtimes 2} (A^{\ot 2*})=
\begin{pmatrix}
x_{11}& x_{12}&x_{21}&x_{22}\\
0 & x_{11} & 0 & x_{21}\\
0 & 0 &x_{11} & x_{12}\\
0 &0&0& x_{11}
\end{pmatrix}
$$
which is $T_{CW,2}$ after permuting basis vectors (see Example \ref{CWqex}) so 
$ GR(W^{\boxtimes 2})=3<4=GR(W)^2$.
\end{example}

 \section{Proofs of main theorems}
 In this section, after reviewing facts about spaces of matrices of bounded rank and 
 the substitution method for bounding tensor rank, we prove 
 a result lower-bounding the tensor rank of tensors associated to compression
 spaces (Proposition \ref{comprex}), a lemma on linear sections of
 $\s_3(Seg(\BP B\times \BP C))$  (Lemma \ref{linelem}), and
 Theorems   \ref{GRtwo}    and \ref{GRthree}.
 
 \subsection{Spaces of matrices of bounded rank}\label{detspaces}
 
Spaces of matrices of bounded rank (Definition \ref{bndrkdef}) is a classical subject dating  back at least to 
  \cite{MR136618}.
The results  
most  relevant here are from
 \cite{MR587090,MR695915}, and they were recast in the language
 of algebraic geometry in \cite{MR954659}. We review notions
 relevant for our discussion.
 
 A large class of spaces of matrices of bounded rank
 $E\subset B\ot C$  are the {\it compression spaces}.
In bases, the space takes the block format
\be\label{comprformat}
 E=\begin{pmatrix} *& * \\ * & 0\end{pmatrix}
\ene
 where if the $0$ is of size  $(\bbb-k)\times (\ccc -\ell)$,
 the space is of bounded rank  $k+\ell$.

  If $m$ is odd, then any linear subspace of $\La 2 \BC^m$ is
  of bounded rank $m-1$.
  More generally one can use the multiplication in any graded algebra
  to obtain   spaces of bounded rank, the   case of $\La 2 \BC^m$ being the
  exterior algebra.

  Spaces of bounded rank at most three  are classified in \cite{MR695915}:
  For three dimensional rank two spaces there are only the compression spaces and 
  the skew symmetric matrices $\La 2\BC^3\subset
  \BC^3\ot \BC^3$.

 \subsection{Review of the substitution method}\label{submethrev}
 \begin{proposition}\cite[Appendix
B]{MR3025382}  \label{prop:slice}
Let $T\in A\ot B\ot C$. 
Fix a basis $a_1\hd a_{\aaa}$ of $A$, with dual basis $\a^1\hd \a^\aaa$.
Write  $T=\sum_{i=1}^\aaa a_i\otimes M_i$, where $M_i =T(\alpha_i)\in B\otimes C$.
Let $\bold R(T)=r$ and $M_1\neq 0$. Then there exist constants $\lambda_2,\dots,
\lambda_\aaa$, such that  the tensor
$$T' :=\sum_{j=2}^\aaa a_j\otimes(M_j-\lambda_j M_1)\in  \tspan\{ a_2\hd a_{\aaa}\}   \ot 
B\ot
C,$$
has rank at most $r-1$.  I.e., $\bold R(T)\geq 1+\bold R(T')$. 

The analogous assertions hold exchanging the role of $A$ with that of $B$ or $C$. 
\end{proposition}

A visual tool for using the substitution method is to write
$T(B^*)$ as a matrix of linear forms. Then the $i$-th row  of  $T(B^*)$ corresponds to a tensor  $a_i\ot M_i\in \BC^1\ot B \ot C$. One adds unknown multiples of the first row of $T(B^*)$ to all other rows, and deletes the first row, then the resulting matrix is $T'(B^*)\in \mathrm{span}\{ a_2\hd a_{\aaa}\}\ot C$. 

In practice one applies   Proposition \ref{prop:slice} iteratively, obtaining a sequence
of tensors in spaces of shrinking dimensions. See    \cite[\S 5.3]{MR3729273}  for a discussion.

 For a positive integer $k\leq\bbb$, if the last $k$ rows of $T(A^*)$ are linearly independent, then one can apply Proposition \ref{prop:slice} $k$ times on the last $k$ rows. In this way, the first $\bbb - k$ rows are modified by unknown linear combinations of the last $k$ rows, and the last $k$ rows are deleted. Then one obtains a tensor $T'\in A\otimes \mathrm{span}\{b_1,\cdots,b_{\bbb-k}\}\otimes C$ such that $\bold R(T')\leq \bold R(T)-k$. 

\begin{proposition}\label{comprex}
Let $T\in A\ot B\ot C$ be a concise tensor with $T(A^*)$ a bounded rank $\rho$
compression space. Then $\bold R(T)\geq \bbb+\ccc-\rho$.  \end{proposition}

\begin{proof} Consider \eqref{comprformat}. Add to the first $k$ rows of $T(A^*)$  unknown linear combinations of the last $\bbb-k$ rows, each of which is 
nonzero by conciseness. Then delete the last $\bbb-k$ rows. Note that the last $\ccc-\ell$ columns
are untouched, and (assuming the most disadvantageous combinations
are chosen) we obtain a tensor   $T'\in A\ot \BC^k\ot C$ 
satisfying $\bold R(T)\geq (\bbb-k)+\bold R(T')$. Next  add to the first $\ell$ columns of $T'(A^*)$ unknown linear combinations of the last $\ccc-\ell$ columns, then delete the last $\ccc-\ell$ columns.
The resulting tensor $T''$ could very well be zero, but 
we nonetheless have $\bold R(T')\geq (\ccc-\ell) +\bold R(T'')$ and thus $\bold R(T)
\geq (\bbb-k)+(\ccc-\ell)=\bbb+\ccc-\rho$.
\end{proof}

Here are the promised lower bounds for $T_{gr3,1deg,m}$:

\begin{example}\label{badgrex}
Consider \eqref{badA}.  Add to  the first row unknown linear combinations of the last $m-1$ rows  then delete the last $m-1$ rows. The resulting tensor is still $\langle a_1\hd a_{q}\rangle$-concise 
so we have $\bold R(T_{gr3,1deg,2q})\geq m-1+ \frac m2$. The case of $T_{gr3,1deg,2q-1}$ is similar.
\end{example}

 \subsection{Lemmas on linear sections of $\s_{r}(Seg(\BP B\times \BP C))$}\label{linelempf}
  
\begin{lemma} \label{turbolinelem}   Let $E\subset B\ot C$
 be a linear subspace.  If $\BP E\cap 
 \s_{r}(Seg(\BP B\times \BP C))$ is a hypersurface in $\BP E$ of degree $r+1$ (counted with multiplicity) and does not contain any hyperplane of $\BP E$, then
 $\BP E \subset \s_{r+1}(Seg(\BP B\times \BP C))$.
 \end{lemma}
 
 \begin{proof}
 Write $E=(y^i_j)$ where $y^i_j=y^i_j(x_1\hd x_{q})$, $1\leq i\leq \bbb$, $1\leq j \leq \ccc$ and $q=\tdim E$. 
 By hypothesis,     all size $r+1$ minors are up to scale equal to a polynomial $S$ of degree $r+1$. No linear 
 polynomial divides $S$ since otherwise the intersection would contain a hyperplane.
Since $\BP E\not\subset \s_{r}(Seg(\BP B\times \BP C))$, there must be a size $r+1$ minor that is nonzero restricted to $\BP E$. Assume it is the $(1\hd r+1)\times (1\hd r+1)$-minor.

\smallskip

 Consider the vector consisting of the first $r+1$ entries of the $(r+2)$-st column. In order that all size $r+1$ minors of the upper left $(r+1)\times(r+2)$ block equal to multiples of $S$, this vector must be a linear combination
of the vectors corresponding to the  first $r+1$ entries of the first $r+1$ columns. By adding linear combinations
of the first $r+1$ columns, we may make these entries zero. Similarly, we may make
all other entries in the first $r+1$ rows zero. By the same argument, we may do 
the same for the first $r+1$ columns.
We have
\be\label{boxformturbo}
\begin{pmatrix} y^1_1 & \cdots  & y^1_{r+1} & 0 & \cdots & 0\\
  \vdots  &  \vdots  &  \vdots  &  \vdots  &  \vdots  & \vdots \\
  y^{r+1}_1 & \cdots  & y^{r+1}_{r+1} & 0 & \cdots & 0\\
  0 & \cdots  & 0 & y^{r+2}_{r+2} & \cdots & y^{r+2}_\ccc\\
  \vdots    & \vdots   & \vdots  & \vdots  &  \vdots  & \vdots  \\
      0 & \cdots  & 0 & y^\bbb_{r+2} & \cdots & y^\bbb_\ccc\end{pmatrix}.
\ene
If $\BP E\not\subset \s_{r+1}(Seg(\BP B\times \BP C))$,  some entry in the
lower $(\bbb-r-1)\times (\ccc-r-1)$ block is nonzero. Take one such and 
  the minor  with it and a $r\times r$ submatrix of
the upper left minor. We obtain a polynomial that has a linear factor, so it cannot be a multiple
of $S$, giving a contradiction.
\end{proof}

\begin{lemma} \label{linelem} Let $\bbb,\ccc>4$. Let $E\subset B
\ot C$
be a linear subspace of dimension $q>4$.  Say $\tdim(\BP E\cap 
\s_{2}(Seg(\BP B\times \BP C))=q-2$ and $\BP E\not\subset \s_{3}(Seg(\BP B\times \BP C))$.
Then either all components of $\BP E\cap 
\s_{2}(Seg(\BP B\times \BP C))$ are   linear  $\pp {q-2}$'s, or $E\subset\BC^5\otimes\BC^5$.
\end{lemma}

The proof is similar to the argument for Lemma \ref{turbolinelem}, except
 that we work in a local ring.

 \begin{proof} 
 
  Write $E=(y^i_j)$ where $y^i_j=y^i_j(x_1\hd x_{q})$, $1\leq i\leq \bbb$, $1\leq j \leq \ccc$. 
 Assume,  to get a contradiction, that there is an irreducible
 component of degree greater than one in the intersection,  given by an irreducible polynomial $S$ of degree
two or three  that divides all size $3$ minors.
By Lemma \ref{turbolinelem},
$\tdeg(S)=2$. 
Since $\BP E\not\subset \s_{2}(Seg(\BP B\times \BP C))$, there must be some size $3$ minor that is nonzero restricted to $\BP E$. Assume it is the $(123)\times (123)$
 minor.

\smallskip
 
 Let $\Delta^I_J$  denote a (signed) size $3$ minor restricted to $E$, where $I=(i_1i_2 i_3)$,
 $J=(j_1j_2 j_3)$, so $\Delta^I_J=L^I_JS$, for some   $L^I_J\in  E^*$.
 Set $I_0=(123)$.
 Consider the $(st4)\times I_0$ minors, where $1\leq s<t\leq 3$.
 Using the Laplace expansion, we may write them as
 \be\label{delta}
 \begin{pmatrix}
 \Delta^{I_0\backslash 1}_{I_0\backslash 1} &\Delta^{I_0\backslash 1}_{I_0\backslash 2} &  \Delta^{I_0\backslash 1}_{I_0\backslash 3}
 \\
\Delta^{I_0\backslash 2}_{I_0\backslash 1} &\Delta^{I_0\backslash 2}_{I_0\backslash 2} &  \Delta^{I_0\backslash 2}_{I_0\backslash 3}
\\
  \Delta^{I_0\backslash 3}_{I_0\backslash 1} & \Delta^{I_0\backslash 3}_{I_0\backslash 2} &  \Delta^{I_0\backslash 3}_{I_0\backslash 3}
 \end{pmatrix}
 \begin{pmatrix}
 y^{4}_1 \\ y^{4}_2 \\ y^{4}_3
 \end{pmatrix}
 =
S \begin{pmatrix}
 L^{234}_{I_0}\\
 L^{134}_{I_0}\\
  L^{124}_{I_0}
 \end{pmatrix}
\ene
Choosing signs properly,  the matrix on the left is just the cofactor matrix of the 
$(123)\times(123)$ submatrix, so its inverse is the transpose
of the original submatrix divided
by the determinant (which is nonzero by hypothesis).
Thus we may write
$$
 \begin{pmatrix}
 y^{4}_1 \\ y^{4}_2 \\ y^{4}_3
 \end{pmatrix}
 =
\frac{S}{\Delta^{I_0}_{I_0}}\begin{pmatrix}
 y^1_1 &  y^2_1 & y^3_1\\
  y^1_2 &  y^2_2 & y^3_2   \\
  y^1_3 &  y^2_3  & y^3_3\end{pmatrix}\begin{pmatrix}
 L^{234}_{I_0}\\
 L^{134}_{I_0}\\
  L^{124}_{I_0}
 \end{pmatrix}.
  $$
 In particular, each $y^{4}_s$, $1\leq s\leq 3$,  is a rational function
 of  $L^{(I_0\backslash t),4}_{I_0}$  and $y^u_v$, $1\leq u,v\leq 3$. 
 $$
 (y^{4}_1, y^4_2, y^{4}_3)=\sum_{t=1}^3 \frac{ L^{(I_0\backslash t),4}_{I_0} }{L^{I_0}_{I_0} }
 (y^{t}_1, y^{t}_2, y^{t}_3).
 $$ 
  Note that the coefficients $\frac{ L^{(I_0\backslash t),4}_{I_0} }{L^{I_0}_{I_0} }$
 are degree zero rational functions in  $L^{(I_0\backslash t),4}_{I_0}$  and $y^u_v$, $1\leq u,v\leq 3$.
  
 The same is true for all $(y^{\ell}_1, y^{\ell}_2, y^{\ell}_3)$ for $\ell\geq 4$.
 Similarly, working with the first $3$ rows  we get
 $(y^{ 1}_\ell, y^{ 2}_\ell, y^{3}_\ell)$ written in terms of
 the $(y_{t}^1, y_{t}^2, y_{t}^3)$ with coefficients  degree zero rational functions in the $y^s_t$.
 Restricting to the Zariski open subset of $E$ where $L^{I_0}_{I_0} \neq 0$,
 we may subtract rational multiples of the first three rows and 
 columns to  normalize our space to \eqref{boxformturbo}:
 \be\label{boxform}
\begin{pmatrix} y^1_1 & y^1_2 & y^1_3 & 0 & \cdots & 0\\
  y^2_1 & y^2_2 & y^2_3 & 0 & \cdots & 0\\
  y^3_1 & y^3_2 & y^3_3 & 0 & \cdots & 0\\
  0 & 0 & 0 & y^4_4 & \cdots & y^4_\ccc\\
  \vdots    & \vdots   & \vdots  & \vdots  &  \vdots  & \vdots  \\
      0 & 0 & 0 & y^\bbb_4 & \cdots & y^\bbb_\ccc\end{pmatrix}.
\ene 
 Since a Zariski open subset of $\BP E$
 is not contained in $\s_2(Seg(\BP B\times \BP C))$, at least one entry
 in the lower right block must be nonzero. Say it is $y^{4}_{4}$.

 On the Zariski open set $L^{I_0}_{I_0}\neq 0$, 
 for all  $ 1\leq s<t\leq 3$, $1\leq u<v\leq 3$,
 we have $y^4_4\Delta^{st}_{uv}=Q^{st}_{uv}S/L^{I_0}_{I_0}$, where $Q^{st}_{uv}$ is a quadratic polynomial (when $\Delta^{st}_{uv}\neq 0$) or zero (when $\Delta^{st}_{uv}=0$). Then either $y^4_4$ is a nonzero multiple of $S/L^{I_0}_{I_0}$, or all $\Delta^{st}_{uv}$'s are multiples of $S$, because $S$ is irreducible.

If all $\Delta^{st}_{uv}$'s are multiples of $S$, 
at least one must be nonzero, say  $\Delta^{12}_{12}\neq 0$. Then by a change of bases we   set $y^3_1,y^3_2,y^1_3,y^2_3$ to   zero. At this point,
for all $1\leq \a,\b\leq 2$,   $\Delta^{\a 3}_{\b 3}$ becomes $y^\a_\b y^3_3$. By hypothesis $\Delta^{\a 3}_{\b 3}$ is a multiple of the irreducible quadratic polynomial $S$, so $y^\a_\b y^3_3=0$. Therefore either all $y^\a_\b$'s are zero, which contradicts with $\Delta^{12}_{12}\neq 0$, or $y^3_3=0$, in which case  all entries in the third row and the third column are zero,   contradicting 
our assumption that the first $3\times 3$ minor is nonzero.

If there exists $\Delta^{st}_{uv}\neq0$ that is not a multiple of $S$, change bases such that it is $\Delta^{12}_{12}$. Note that $y^4_4=\Delta^{1234}_{1234}/\Delta^{123}_{123}=(\Delta^{1234}_{1234}/S)/L^{I_0}_{I_0}$ where $(\Delta^{1234}_{1234}/S)$ is a quadratic polynomial. By hypothesis $S$ divides $\Delta^{124}_{124}=\Delta^{12}_{12}y^4_4$. Since $S$ is irreducible and $\Delta^{12}_{12}$ is not a multiple of $S$, $\Delta^{1234}_{1234}/S$ must be a multiple of $S$. Therefore  $\Delta^{1234}_{1234}$ is a multiple of $S^2$. This is true for all size $4$ minors, therefore we can apply \ref{turbolinelem}. By the proof of \ref{turbolinelem}, all entries of $E$ can be set to  zero except those in the upper left $5\times 5$ block, so $E\subset\BC^5\otimes\BC^5$.
\end{proof}
 
 \begin{remark}  
  The normalization in the case $\tdeg(S)=2$  is not possible in general without the restriction
  to the open subset where $L^{I_0}_{I_0}\neq 0$. Consider 
 $
 T(A^*)
 $
 such that the upper $3\times 3$ block
 is
 $$\begin{pmatrix} x_1& x_2&x_3\\ -x_2 & x_1 & x_4\\
 -x_3& -x_4& x_1\end{pmatrix}.
 $$
 Then the possible entries in the first three
 columns of the fourth row are not limited to the span of
 the first three rows. The element
 $(x_4,-x_3,x_2)$ is also possible.
 \end{remark}

\subsection{Proof of Theorem \ref{GRtwo}}\label{GRtwopf}

We recall the statement:

\begin{customthm} {\ref{GRtwo}}  For $\aaa,\bbb,\ccc\geq3$, 
the variety $\mathcal{GR}_{2}(A\ot B\ot C)$ is the variety of tensors $T$ such that $T(A^*)$, $T(B^*)$,
 or $T(C^*)$ has bounded rank 2. 
 
There are exactly two concise tensors in $\mathbb{C}^3\otimes\mathbb{C}^3\otimes\mathbb{C}^3$ with $GR(T)=2$:
\begin{enumerate}
\item The unique up to scale skew-symmetric tensor $T=\sum_{\s\in \FS_3}\tsgn(\s) a_{\s(1)}\ot b_{\s(2)}\ot c_{\s(3)}\in \La 3\BC^3\subset  \BC^3\ot \BC^3\ot \BC^3$

and

\item $T_{utriv,3}:=a_1\ot b_1\ot c_1 + a_1\ot b_2\ot c_2+ a_1\ot b_3\ot c_3
+a_2\ot b_1\ot c_2+a_3\ot b_1\ot c_3\in S^2\BC^3\ot \BC^3\subset  \BC^3\ot \BC^3\ot \BC^3$. 
\end{enumerate}
There is a unique  concise  tensor $T\in  \BC^m\ot \BC^m\ot \BC^m$
satsifying $GR(T)=2$ when $m>3$,    
  namely
 $$
 T_{utriv,m}:=
 a_1\ot b_1\ot c_1+   \sum_{\rho=2}^m [a_1\ot  b_\rho\ot c_\rho +   a_\rho\ot b_1\ot c_\rho].
 $$
 This tensor satisfies $\ur(T_{utriv,m})=m$ and $\bold R(T_{utriv,m})=2m-1$.  
\end{customthm}
 
 \begin{proof} For simplicity, assume $\aaa\leq \bbb\leq \ccc$. 
A tensor $T\in A\ot B\ot C$ has geometric rank 2 if and only if
$\mathbb{P}T(A^*)\not\subset Seg(\BP B\times \BP C)$, 
and  either $\mathbb{P}T(A^*)\cap Seg(\BP B\times \BP C)$ has dimension $\aaa-2$  or $\mathbb{P}T(A^*)\subset\sigma_2( Seg(\BP B\times \BP C))$.
  
 For the case $\mathbb{P}T(A^*)\subset\sigma_2( Seg(\BP B\times \BP C))$, $T(A^*)$ is of bounded rank $2$. By the classification of spaces of bounded rank $2$, up to equivalence $T(A^*)$ must be in one of the following forms:
\begin{center}
$\begin{pmatrix}
* & \cdots & * \\
* & \cdots & * \\
0 & \cdots & 0 \\
\vdots & &\vdots\\
0 & \cdots & 0
\end{pmatrix}$,
$\begin{pmatrix}
* & * &\cdots & * \\
* & 0 &\cdots & 0 \\
\vdots &\vdots & &\vdots\\
* & 0 &\cdots & 0
\end{pmatrix}$,  or 
$\begin{pmatrix}
0&x&y& 0 &\cdots & 0\\
-x&0&z& 0 &\cdots & 0\\
-y&-z&0& 0 &\cdots & 0\\
0 &&&\cdots && 0\\
\vdots   &&&\vdots &&  \vdots\\
0 &&&\cdots && 0 
\end{pmatrix}$.
\end{center}

When $T$ is concise, it must be of the second form or the third
with $\aaa=\bbb=\ccc=m=3$. If it is the third, $T$ is the unique up to scale skew-symmetric tensor in $ \mathbb{C}^3\ot \BC^3\ot \BC^3$. If it is of the second form and $\aaa=\bbb=\ccc=m$, the entries in the first column   must be linearly independent, as well as the entries in the first row. Thus we may choose a basis of $A$ such that $T=a_1\otimes b_1\otimes c_1 + \sum^m_{i>1} y_i\otimes b_i\otimes c_1 + \sum^m_{i>1}z_i\otimes b_1 \otimes c_i$ where $y_i$'s and $z_i$'s are linear combinations of $a_2,\cdots,a_m$. Then by a change of basis in $b_2,\cdots,b_m$ and $c_2,\cdots,c_m$ respectively, we obtain $T_{utriv,m}$.

 For the case $\tdim(\mathbb{P}T(A^*)\cap Seg(\BP B\times \BP C))=\aaa-2$, by Lemma \ref{turbolinelem}, if this intersection is an irreducible quadric, we are
 reduced to the case
 $\mathbb{P}T(A^*)\subset\sigma_2( Seg(\BP B\times \BP C))$.
 Thus all $2\times2$ minors of $T(A^*)$  have   a common linear factor.  
Assume 
the common factor  is $x_1$. Then $\mathbb{P}T(A^*)\cap Seg(\BP B\times \BP C)\supset\{x_1=0\}$. Hence $\mathbb{P}T(\langle\alpha_2,\cdots,\alpha_\aaa\rangle)\subset Seg(\BP B\times \BP C)$, i.e. $T(\langle\alpha_2,\cdots,\alpha_\aaa\rangle)$ is of bounded rank one. By a change of bases in $B,C$ and exchanging    $B$ and $C$, all entries but the first row of $T(\langle\alpha_2,\cdots,\alpha_\aaa\rangle)$ becomes zero. Then all entries but the first column and the first row of $T(C^*)$ are zero, so $T(C^*)$ is of bounded rank $2$.

When $T$ is concise and $\aaa=\bbb=\ccc=m$,   the   change of bases as in the case when $T(A^*)$ is of bounded rank $2$
shows that  up to a reordering of   $A$, $B$ and $C$ we obtain $T_{utriv,m}$.
\end{proof}

 \subsection{Proof of Theorem \ref{GRthree}}\label{GRthreepf}
 We recall the statement:
 \begin{customthm}{\ref{GRthree} } Let $T\in A\ot B\ot C$ be concise and assume $\ccc\geq\bbb\geq \aaa>4$.
If $GR(T)\leq 3$, then  
$\bold R(T)\geq  \bbb+ \lceil {\frac {\aaa-1}2}\rceil -2$.

If moreover $\aaa=\bbb=\ccc=m$ and $T$ is 
  $1_*$-generic, then $\bold R(T)\geq 2m-3$. 
\end{customthm}
\begin{proof}  
  In order for $GR(T)=3$,   either $\BP T(A^*)\subset \s_3(Seg(\BP B\times \BP C))$,
   $\BP T(A^*)\cap \s_2(Seg(\BP B\times \BP C))$ has dimension $\aaa-2$, 
  or $\BP T(A^*)\cap Seg(\BP B\times \BP C)$ has dimension $\aaa-3$.
 
Case  $\BP T(A^*)\subset \s_3(Seg(\BP B\times \BP C))$: Since $\aaa>3$,
it must be a compression space. We conclude by Proposition  \ref{comprex}.

\medskip

Case
 $\BP T(A^*)\cap \s_2(Seg(\BP B\times \BP C))$ has dimension $\aaa-2$: 
By  Lemma  \ref{linelem},   there exists $a\in A$
such that $\BP T( a^\perp)\subset  \s_2(Seg(\BP B\times \BP C))$.
Write $T(A^*)=x_1Z+U$, where $Z$ is a matrix of scalars
and $U=U(x_2\hd x_\aaa)$ is a matrix of linear forms of bounded rank two.
As discussed in \S\ref{detspaces}, there are two possible normal forms for $U$ up  to symmetry.

If $U$ is zero outside of the first two rows, 
add to the first two rows an unknown combination of the last $\bbb-2$
rows (each of which is nonzero by conciseness), so that the resulting tensor $T'$ satisfies
$\bold R(T)\geq \bbb-2+\bold R(T')$. Now the last $\bbb-2$ rows only contained
multiples of $x_1$ so $T'$ restricted to $a_1^\perp$ is $\langle a_2\hd a_\aaa\rangle$-concise 
and thus of rank at least $\aaa-1$,  so $\bold R(T)\geq \aaa+\bbb-3$.

Now say  $U$ is zero outside its first row and column.

Subcase: $\aaa=\bbb=\ccc$ and $T$ is $1_*$-generic. Then either $T$ is $1_A$-generic, or the first
row or column of $U$ consists of linearly independent entries.
If  $T$ is $1_A$-generic, we may change bases so that  $Z$ is of full rank.
Consider $T(B^*)$. Its first row consists
of linearly independent entries.  Write $\aaa=\bbb=\ccc=:m$ and  apply the substitution method to  delete the last $m-1$ rows  (each
of which is nonzero by conciseness). Call the resulting tensor
$T'$, so $\bold R(T)\geq \bold R(T')+m-1$. 
 Let $T''=T'|_{A^*\otimes\mathrm{span}\{\b_2,\cdots,\b_m\} \otimes\mathrm{span}\{\g_2,\cdots,\g_m\} }$.  Then $T''(A^*)$ equals to the matrix obtained by removing the first column and the first row from $x_1Z$, so $\bold R(T'')\geq \bold R(x_1Z)-2=m-2$.  Thus $\bold R(T)\geq (m-1)+m-2$ and we conclude.
If the first row of $U$ consists of linearly independent entries, then  the
same argument, using $T(A^*)$,  gives the bound.

Subcase: $T$ is $1$-degenerate or $\aaa,\bbb,\ccc$ are not all equal.
By $A$-conciseness, either  the first row or column
 must have at least $\lceil\frac{\aaa-1 }2\rceil$ independent entries of $\mathrm{span}\{x_2,\cdots,x_{\aaa}\}$.  Say it is the 
first row.
Then applying the substitution method, adding an unknown combination
of the last $\bbb-1$ rows to the first, then deleting the last $\bbb-1$ rows. Note that all entries in the first
row except  the $(1,1)$ entry are only
altered by multiples of $x_1$,  so there are at least $\lceil\frac{\aaa-1 }2\rceil -1$ linearly independent entries in the resulting matrix.
We obtain 
 $\bold R(T)\geq \bbb-1+ \lceil\frac{\aaa-1 }2\rceil -1$.

\medskip

Case 
  $\tdim(\BP T(A^*)\cap Seg(\BP B\times \BP C))=\aaa-3$:
We split this into three sub-cases based on the dimension
of the span of the intersection:
\begin{enumerate}
\item $\tdim \langle\BP T(A^*)\cap Seg(\BP B\times \BP C)\rangle =\aaa-3$
\item $\tdim \langle\BP T(A^*)\cap Seg(\BP B\times \BP C)\rangle =\aaa-2$
\item $\tdim \langle\BP T(A^*)\cap Seg(\BP B\times \BP C)\rangle =\aaa-1$
\end{enumerate}

Sub-case (1): the intersection must be a linear space.
We may choose bases such that 
$$
T(A^*)= 
\begin{pmatrix} 
0 & 0 & x_3& \cdots & x_\aaa\\
0& 0& 0 & \cdots & 0\\
& & \ddots & & \\
& & & \ddots & \\
& & & & 0\end{pmatrix} + x_1Z_1+ x_2Z_2
$$
where $Z_1,Z_2$ are  $\bbb\times \ccc$ scalar matrices.
Add to  the first row   a linear combination of   the last $\bbb-1$ rows
(each of which is nonzero by conciseness)
to obtain a tensor of rank at least $\bbb-2$, giving   $\bold R(T)\geq \aaa+\bbb-3$.

Sub-case (2): By Lemma \ref{turbolinelem}   the intersection must 
contain a $\pp{\aaa-2}$, and one argues
as in case (1), except there is just $x_1Z_1$ and $x_2$ also appears in the first row.
 
\medskip
Sub-case (3):  
$T(A^*)$ must have a basis of elements of rank one.
The only way $T$ can be concise, is for $\aaa=\bbb=\ccc=m$ and   $m$ elements
of the $B$ factor form a basis and same for the $C$
factor. Changing bases, we
have $T=\sum_{j=1}^ma_j\ot b_j\ot c_j$ which just intersects
the Segre in points, so this case cannot occur.
 \end{proof}

  \bibliographystyle{amsplain}

\bibliography{Lmatrix}

 \end{document}